\documentclass{article}
\usepackage{amscd,amsmath,amssymb,bbm}
\usepackage{amsfonts}
\newcommand{\p}[1]{(\ref{#1})}

\newcommand{\be}{\begin{equation}}
\newcommand{\ee}{\end{equation}}
\newcommand{\bea}{\begin{eqnarray}}
\newcommand{\eea}{\end{eqnarray}}
\newcommand{\ba}{\begin{array}}
\newcommand{\ea}{\end{array}}

\newcommand{\und}{\qquad\textrm{and}\qquad}

\newcommand{\nn}{\nonumber}

\newcommand{\R}{{\mathbb{R}}}

\def\di{{\rm d}}
\def\im{{\rm i}}
\def\ep{{\rm e}}
\def\={\ =\ }

\def\Nt{$\cal N${=\,}2~}
\def\Nf{$\cal N${=\,}4~}
\def\Ne{$\cal N${=\,}8~}

\begin{document}
\thispagestyle{empty}
\vspace{2cm}
\begin{flushright}
\end{flushright}\vspace{2cm}
\begin{center}
{\LARGE\bf SU(2) reductions in \Nf multidimensional supersymmetric mechanics}
\end{center}
\vspace{1cm}

\begin{center}
{\Large
Stefano Bellucci$\,{}^{a}$, Sergey Krivonos$\,{}^{b}$  and
Anton Sutulin$\,{}^{b}$
}\\
\vspace{1.0cm}
${}^a$ {\it
INFN-Laboratori Nazionali di Frascati,
Via E. Fermi 40, 00044 Frascati, Italy} \vspace{0.2cm}

${}^b$
{\it Bogoliubov  Laboratory of Theoretical Physics,
JINR, 141980 Dubna, Russia}
\vspace{0.2cm}
\end{center}
\vspace{3cm}

\begin{abstract}
\noindent We perform an $su(2)$ Hamiltonian reduction in the
bosonic sector of the  $su(2)$-invariant action for two free
$(4,4,0)$ supermultiplets. As a result, we get the  five
dimensional \Nf supersymmetric mechanics describing the motion of an
 isospin carrying particle interacting with a Yang
monopole. We provide the Lagrangian and Hamiltonian descriptions
of this system.
Some possible generalizations of the action to
the cases of systems with a more general bosonic action, a
four-dimensional system which still includes eight fermionic
components, and a variant of five-dimensional \Nf mechanics
constructed with the help of the ordinary and twisted \Nf
hypermultiplets were also considered.
\end{abstract}

\newpage
\setcounter{page}{1}

\setcounter{equation}0
\section{Introduction}
The supersymmetric mechanics describing the motion of an isospin
particle in the background non-Abelian gauge fields attracted a
lot of attention in the last few years \cite{brazil,Kon1, Kon2,
dual, fil0, Kon3, bk, sutulin1, sutulin2, KL1}, especially due to
its close relation with higher dimensional Hall effects and their
extensions \cite{HE}, as well as with the supersymmetric versions
of various Hops maps (see e.g. \cite{brazil}). The key point of
any possible construction is to find proper ``room''  for
semi-dynamical ``isospin'' variables, which have to be invented
for the description of monopole-type interactions in Lagrangian
mechanics. In supersymmetric systems these isospin variables
should belong to some supermultiplet, and the main question is
what to do with additional fermions accompanying the isospin
variables. In \cite{Kon2} the additional fermions, together with
isospin variables, span an auxiliary $(4,4,0)$ multiplet with a
Wess-Zumino type action possessing an extra $U(1)$ gauge
symmetry\footnote{Note, that the first realization of this idea
was proposed in \cite{fil0}}. In this framework, an off-shell
Lagrangian formulation was constructed, with the harmonic
superspace approach \cite{hss,NLM0}, for a particular class of
four-dimensional \cite{Kon2} and three-dimensional \cite{Kon3} \Nf
mechanics, with self-dual non-Abelian background. The same idea of
coupling with an auxiliary ``semi-dynamical'' supermultiplet has
been also elaborated in \cite{bk}, within the standard \Nf
superspace framework, and then it has been applied for the
construction of the Lagrangian and Hamiltonian formulations of the
\Nf supersymmetric system describing the motion of the  isospin
particles in three \cite{sutulin1} and four-dimensional
\cite{sutulin2} conformally flat manifolds, carrying the
non-Abelian fields of the Wu-Yang monopole and BPST instanton,
respectively.

In both these approaches the additional fermions were completely
auxiliary, and they were expressed through the physical ones on
the mass shell. Another approach based on the direct use of the
$SU(2)$ reduction, firstly considered on the Lagrangian level in
the purely bosonic case in \cite{brazil}, has been used in supersymmetric case in
\cite{KL1}. The key idea of this approach is to perform a direct
$su(2)$ Hamiltonian reduction in the bosonic sector of the \Nf
supersymmetric system, with the general $SU(2)$ invariant action
for a self-coupled $(4,4,0)$ supermultiplet. No auxiliary
superfields are needed within such an approach, and the procedure
itself is remarkably simple and automatically successful.

As concerning the interaction with the non-Abelian background, the
system considered  in \cite{KL1} was not too illuminating, due to
its small number (only one) of physical bosons. In the present
Letter we extend the construction of \cite{KL1} to the case of the
\Nf supersymmetric system with five (and four in the special case)
physical bosonic components. It is not, therefore, strange that
the arising non-Abelian background coincides with the field of a
Yang monopole (a BPST instanton field in the four dimensional
case). The very important preliminary step, we discussed in
details in Section 2, is to pass to new bosonic and fermionic
variables, which are inert under the $SU(2)$ group, over which we
perform the reduction. Thus, the $SU(2)$ group rotates only the
three bosonic components, which enter the action through $SU(2)$
invariant currents. Just these bosonic fields become  the
``isospin'' variables which the background field couples to. Due
to the commutativity of \Nf supersymmetry with the reduction
$SU(2)$ group, it survives upon reduction. In Section 3 we present
the corresponding supercharges and Hamiltonian, which form a
standard \Nf superalgebra. In Section 4 we consider some possible
generalizations, which include a system with more general bosonic
action, a four-dimensional system which still includes eight
fermionic components, and the variant of five-dimensional \Nf
mechanics constructed with the help of ordinary and twisted \Nf
hypermultiplets. Finally, in the Conclusion we discuss some
unsolved problems and possible extensions of the present
construction.

\setcounter{equation}0
\section{$(8_B,8_F)\rightarrow (5_B,8_F)$ reduction and Yang monopole}
As the first nontrivial example of the $SU(2)$ reduction in \Nf
supersymmetric mechanics we consider the reduction from the
eight-dimensional bosonic manifold to the five dimensional one. To
start with, let us choose our basic \Nf superfields to be the two
quartets of real \Nf superfields ${\cal Q}^{i\hat\alpha}_A$ (with
$i,\hat\alpha,A=1,2$) defined in the \Nf superspace
$\R^{(1|4)}=(t,\theta_{ia})$ and subjected to the constraints
\be\label{con}
D^{(ia}{\cal Q}^{j)\hat\alpha}_A=0\ ,  \und \left(
{\cal Q}^{i\hat\alpha}_A\right)^\dagger = {\cal Q}_{i\hat\alpha
A}\ ,
\ee
where the corresponding covariant derivatives have the
form
\be
D^{ia}=\frac{\partial}{\partial
\theta_{ia}}+\im\theta^{ia} \partial_t\ , \qquad\textrm{so
that}\qquad \bigl\{ D^{ia}, D^{jb}\bigr\}=2i
\epsilon^{ij}\epsilon^{ab}\partial_t\ .
\ee
Each of these \Nf
supermultiplets describes four bosonic and four fermionic variables
off-shell~ \cite{NLM0,hyper1,hyper2,hyper3,hyper4,ikl1}.

The most general action for ${\cal Q}^{i\hat\alpha}_A$ superfields
is constructed by integrating an arbitrary superfunction ${\cal
F}({\cal Q}^{i\hat\alpha}_A)$ over the whole \Nf superspace. Here,
we restrict ourselves to the simplest prepotential of the
form\footnote{We used the following definition of the superspace
measure: $\di^4\theta \equiv
-\frac{1}{96}D^{ia}D_{ib}D^{bj}D_{ja}$.}
\be\label{action1}
{\cal
F}({\cal Q}^{i\hat\alpha}_A)\={\cal Q}^{i\hat\alpha}_A {\cal
Q}_{i\hat\alpha A} \qquad\longrightarrow\qquad
S\=\int\!\!\di{t}\,\di^4\theta\ {\cal Q}^{i\hat\alpha}_A {\cal
Q}_{i\hat\alpha A}.
\ee
The rationale for this selection is,
first of all, its manifest invariance under $su(2)$
transformations acting on the ``$\hat\alpha$'' index of~${\cal
Q}{}^{i\hat\alpha}$. This is the symmetry over which we are going
to perform the $su(2)$ reduction. Secondly, just this form of the
prepotential guarantees $SO(5)$ symmetry in the  bosonic sector
after reduction.

In terms of components the action \p{action1}  reads
\be\label{action2}
S\=\int\!\!\di{t}\left[ {\dot Q}{}^{i\hat\alpha}_A\; {\dot Q}{}_{i\hat\alpha A} -\frac{i}{8}{\dot \Psi}{}^{a\hat\alpha}_A \Psi_{a\hat\alpha A}\right]
\ee
where the bosonic and fermionic components are defined as
\be\label{comp1}
Q^{i\hat\alpha}_A={\cal Q}^{i\hat\alpha}_A|\,, \qquad \Psi{}^{a\hat\alpha}_A= D^{ia}{\cal Q}^{\hat\alpha}_{iA}|\,,
\ee
and, as usually, $(\ldots)|$ denotes the $\theta_{ia}=0$ limit.
Thus, from beginning we have just the sum of two independent non-interacting
$(4,4,0)$ supermultiplets.

To proceed further we define the following bosonic $q_A^{i\alpha}$
and fermionic $\psi^{a\alpha}_A$ fields
\be\label{comp2}
q_A^{i\alpha} \equiv Q^i_{A\hat\alpha} G^{\alpha\hat\alpha}, \qquad
\psi^{a\alpha}_A \equiv \Psi^{a}_{\hat\alpha A}G^{\alpha\hat\alpha},
\ee
where  the  bosonic variables $G^{\alpha\hat\alpha}$, subjected
to $G^{\alpha\hat\alpha }G_{\alpha\hat\alpha}=2$, are chosen as
\be\label{q}
G^{11}= \frac{\ep^{-\frac{\im}2\phi}}{\sqrt{1{+}\Lambda\overline\Lambda}}\,\Lambda
\ ,\qquad
G^{21}=-\frac{\ep^{-\frac{\im}2\phi}}{\sqrt{1{+}\Lambda\overline\Lambda}}
\ , \qquad
G^{22}= \left(G^{11}\right)^\dagger\ ,\qquad
G^{12}=-\left(G^{21}\right)^\dagger\ .
\ee
The variables $G^{\alpha\hat\alpha}$ play the role of bridge relating two different $SU(2)$ groups
realized on the indices $\alpha$ and $\hat\alpha$ respectively.
Thus, we split our eight bosonic variables $Q^i_{A\hat\alpha}$ \p{comp1} into
five ones in $q_A^{i\alpha}$ and three fields $(\phi,\Lambda, \overline \Lambda)$
entering via $G^{\alpha\hat\alpha}$ \p{q}.

Now our action \p{action2} acquires the form
\be\label{action3}
S\=\int\!\!\di{t}\left[ {\dot q}{}^{i\alpha}_A\; {\dot q}{}_{i\alpha A}-2 q^{i\alpha}_A{\dot q}{}^\beta_{iA}J_{\alpha\beta}+
\frac{q^{i\alpha}_A q_{i\alpha A}}{2} J^{\beta\gamma}J_{\beta\gamma} -\frac{i}{8}{\dot \psi}{}^{a\alpha}_A \psi_{a\alpha A} + \frac{i}{8}\psi_A^{a\alpha}\psi_{aA}^\beta J_{\alpha\beta} \right],
\ee
where
\be\label{J}
J^{\alpha\beta} =J^{\beta\alpha} = G^{\alpha\hat\alpha}{\dot G}{}^\beta_{\hat\alpha}.
\ee

The new variables $q_A^{i\alpha}$ and $\psi^{a\alpha}_A$, which,
clearly, contain  five independent bosonic and eight fermionic
components, are inert under $su(2)$ rotations of the hatted
indices. Under these $su(2)$ rotations, realized now only on
$G^{\alpha\hat\alpha}$ variables in a standard way
$$
\delta G^{\alpha\hat\alpha}=\gamma^{(\hat\alpha\hat\beta)}G^\alpha_{\hat\beta},
$$
the fields $(\phi,\Lambda,\bar\Lambda)$ \p{q} transform as ~\cite{ikl1}
\be\label{transf1}
\delta\Lambda = \gamma^{11}\ep^{\im\phi}(1{+}\Lambda\overline\Lambda)\ ,\qquad
\delta\overline\Lambda = \gamma^{22}\ep^{-\im\phi}(1{+}\Lambda\overline\Lambda)\ ,\qquad
\delta\phi=-2\im\gamma^{12}+
\im\gamma^{22}\ep^{-\im\phi}\Lambda -\im\gamma^{11}\ep^{\im\phi}\overline\Lambda\ .
\ee
It is easy to check that the forms $J^{\alpha\beta}$ \p{J} entering the action \p{action3} and having the following
explicit form
\be\label{omega}
J^{11}\=-\frac{\dot\Lambda- i \Lambda\dot\phi}{1+\Lambda\overline\Lambda}
\ ,\qquad
J^{22}\=-\frac{\dot{\overline\Lambda}+ i \overline\Lambda\dot\phi}{1+\Lambda\overline\Lambda} \und
J^{12}\=- i \frac{1-\Lambda\overline\Lambda}{1+\Lambda\overline\Lambda}\,\dot\phi\ -\
\frac{\dot\Lambda\overline\Lambda-\Lambda\dot{\overline\Lambda}}{1+\Lambda\overline\Lambda}
\ee
are also invariant under~\p{transf1}, as is the whole action \p{action3}.

Next, we introduce the standard Poisson brackets \be\label{pb}
\left\{\pi,\Lambda\right\}=1\ ,\qquad
\left\{\bar\pi,\overline\Lambda\right\}=1\ ,\qquad
\left\{p_{\phi},\phi\right\}=1\ , \ee so that the generators of
the transformations \p{transf1}, \be\label{currents}
I_{\phi}\=p_{\phi}\ ,\qquad
I\=\ep^{\im\phi}\bigl[(1{+}\Lambda\overline\Lambda)\,\pi-\im\overline\Lambda\,p_\phi\bigr] \ ,\qquad
{\bar I}\=\ep^{-\im\phi}\bigl[(1{+}\Lambda\overline\Lambda)\,\bar\pi+
\im\Lambda\,p_\phi\bigr]\ ,
\ee
will be the Noether constants of
motion for the action~\p{action3}. To perform the reduction over
this SU(2) group we fix the Noether constants
as~(c.f.~\cite{brazil})
\be\label{reduction1}
I_\phi=m \und
I={\bar I}=0\ ,
\ee
which yields
\be\label{reduction2}
p_\phi\=m
\und \pi\=\frac{\im m\,\overline\Lambda}{1+\Lambda\overline\Lambda}\ ,\qquad
\bar\pi\=-\frac{\im m\,\Lambda}{1+\Lambda\overline\Lambda}\ .
\ee
Performing a Routh transformation over the variables
$(\Lambda, \overline\Lambda, \phi)$, we reduce the action~\p{action3} to
\be\label{action4}
{\widetilde S}\ \=\ S\ -\ \int\!\!\di{t}\
\bigl\{
\pi\,{\dot\Lambda}+\bar\pi\,\dot{\overline\Lambda}+p_\phi{\dot\phi}\bigr\}
\ee
and substitute the expressions~\p{reduction2} into~$\tilde S$.

As the final step we have to choose the proper parametrization for
bosonic components $q_A^{i\alpha}$ \p{comp2} remembering that they
contain only five independent variables. Following \cite{brazil}
we will choose these variables as
\be\label{eq1}
q_1^{i\alpha}=\frac{1}{2}\epsilon^{i\alpha} \sqrt{r+z_5},\quad
q_2^{i\alpha}=\frac{1}{\sqrt{2(r+z_5)}}\left(
x^{(i\alpha)}-\frac{1}{\sqrt{2}}\epsilon^{i\alpha}z_4\right),
\ee
where
\be\label{coord} x^{12}=\frac{i}{\sqrt{2}}z_3,
x^{11}=\frac{1}{\sqrt{2}}\left( z_1+i z_2\right),
x^{22}=\frac{1}{\sqrt{2}}\left( z_1-i z_2\right), \quad \mbox{and}
\quad r^2=\sum_{i=1}^5 z_i z_i ,
\ee
and five independent fields
are $z_m, m=1,...,5$. A slightly lengthy but straightforward
calculations give the action
\bea\label{action5}
S_{\rm red}&{=}&
\int\!\!\di{t}\left[ \frac{1}{4r} {\dot z_m} {\dot z_m} -
\frac{i}{8}{\dot\psi}{}^{a\alpha}_A \psi_{a\alpha
A}+\frac{i}{4r}H^{\alpha\beta}V_{\alpha\beta}+
\frac{1}{128 r} H^{\alpha\beta}H_{\alpha\beta}\right. \nn \\
& &\left. -\frac{m^2}{r}  -\frac{m}{4r} v^\alpha {\bar v}{}^\beta
H_{\alpha\beta}-\frac{4im}{r} v^\alpha{\bar v}{}^\beta
V_{\alpha\beta}+ im\left( {\dot v}{}^\alpha{\bar v}_\alpha
-v^\alpha\dot{\bar v}_\alpha\right) \right]. \eea Here
\be\label{H} H^{\alpha\beta}=\psi^{a\alpha}_A \psi^\beta_{aA},
\qquad  v^\alpha=G^{\alpha 1},\; {\bar v}{}^\alpha = G^{\alpha
2},\; v^\alpha{\bar v}_\alpha=1, \ee and to ensure that the
reduction constraints \p{reduction2} are satisfied we added
Lagrange multiplier terms (last two terms in \p{action5}).
Finally, the variables $V^{\alpha\beta}$ entering the action
\p{action5} are  defined in a rather symmetric way to be
\be\label{ym} V^{\alpha\beta}=\frac{1}{2}\left( q^{i\alpha}_A
{\dot q}^\beta_{iA}+q^{i\beta}_A {\dot q}^\alpha_{iA} \right). \ee
To clarify the relations of these variables with the potential of
Yang monopole, one has to introduce the following isospin currents
(which will form $su(2)$ algebra upon quantization): \be\label{T}
T^I=v^\alpha \left( \sigma^I \right)_\alpha^\beta {\bar v}_\beta,
\qquad I=1,2,3. \ee Now, the ``harmonics'' ($v^\alpha {\bar
v}{}^\beta$)-dependent terms in the action \p{action5} can be
rewritten as \be\label{Y} -\frac{m}{4r} v^\alpha {\bar v}{}^\beta
H_{\alpha\beta} -\frac{4im}{r} v^\alpha{\bar v}{}^\beta
V_{\alpha\beta} = m\; T^I\;\left( \frac{1}{8r}H^I +
\frac{1}{r(r+z_5)}\eta^I_{\mu\nu} z_\mu {\dot z}_\nu \right), \;
\mu,\nu=1,2,3,4. \ee Here, we defined the fermionic spin currents
\be H^I=H^\alpha_\beta \left( \sigma^I \right)_\alpha^\beta \ee
and a self-dual t'Hooft symbol \be \eta^I_{\mu\nu}=\delta^I_\mu
\delta_{\nu 4}-\delta^I_\nu \delta_{\mu 4}+\epsilon^A{}_{\mu\nu
4}. \ee Thus we conclude, that the action \p{action5} describes
\Nf supersymmetric five-dimensional isospin particles moving in
the field of Yang monopole \be {\cal A}_\mu=
-\frac{1}{r(r+z_5)}\eta^I_{\mu\nu} z_\nu T^I . \ee We stress that
the $su(2)$ reduction algebra, realized in~\p{transf1}, commutes
with all (super)symmetries of the action~\p{action2}. Therefore,
all symmetry properties of the theory are preserved in our
reduction and the final action \p{action5} possesses \Nf
supersymmetry.

\setcounter{equation}0
\section{Hamiltonian and Supercharges}
Within our approach the construction of the supercharges is
straightforward. To simplify the expressions for the supercharges
in this Section we will choose a slightly different
parametrization of the physical components as compared to \p{eq1}.
Namely, we will choose the physical bosonic components as follows:
\be\label{1a}
q_1{}^{i\alpha}=\frac{1}{\sqrt{2}}\epsilon^{i\alpha}
e^{\frac{1}{2}u},\quad q_2{}^{i\alpha}.
\ee
With this
parametrization the reduced action \p{action5} reads
\bea\label{2a}
S_{\rm red}&=& \int dt \left[ \frac{1}{4}e^u {\dot
u}{}^2+ \frac{e^u}{e^u+y}({\dot q}_2\cdot {\dot q}_2)+
\frac{1}{e^u+y}(q_2 \cdot {\dot q}_2)^2 +\frac{1}{128(e^u+y)}(H\cdot H)+\frac{\im}{4(e^u+y)}(V\cdot H) \right.\nn\\
& &  -\frac{m^2}{e^u+y}-\frac{m}{4(e^u+y)}v^{\alpha}{\bar v}{}^\beta H_{\alpha\beta}-
\frac{4 \im m}{e^u+y}v^{\alpha}{\bar v}{}^\beta V_{\alpha\beta} \nn \\
&& \left. -
\frac{i}{8}\left({\dot\psi}_1\cdot \psi_{1}+{\dot\psi}_2 \cdot \psi_{2}\right)+
im\left( {\dot v}{}^\alpha{\bar v}_\alpha -v^\alpha\dot{\bar v}_\alpha\right) \right].
\eea
Here,
\be\label{3a}
y= q_2{}^{i\alpha}q_{2\; i\alpha},
\ee
while the currents $J^{\alpha\beta}, V^{\alpha\beta}, H^{\alpha\beta}$ and the ``harmonics'' $v^{\alpha},{\bar v}_\alpha$ are defined
as previously  in \p{J}, \p{H} and \p{ym}.

One may check that the action \p{2a} is invariant under \Nf supersymmetry transformations (with parameters $\mu^{ia}$)
\bea\label{4a}
\delta u &=& -\frac{1}{\sqrt{2}}e^{-\frac{1}{2}u} \mu^{ia}\psi_{1\; a i}, \quad
\delta G^{i\hat\alpha}= \frac{1}{2\sqrt{2}}e^{-\frac{1}{2}u}\left( \mu^{jb}\psi_{1\; bj}G^{i\hat\alpha}-
2\mu^{ib}\psi_{1\; bj}G^{j\hat\alpha}\right),\nn \\
\delta \psi_1^{a\alpha}&=& -\im \sqrt{2} \mu^{\alpha a}e^{\frac{1}{2}u}{\dot u}-\im 2\sqrt{2}e^{\frac{1}{2}u}\mu^{\beta a}J_\beta{}^\alpha+
\frac{e^{-\frac{1}{2}u}}{2\sqrt{2}}
\left(\mu^{jb}\psi_{1\;bj}\psi_1^{a\alpha}-2\mu^{\alpha b}\psi_{1\;bj}\psi_1^{aj}\right),\nn\\
\delta q_2^{i\alpha}&=& \frac{1}{2}\mu^{ia}\psi_{2\;a}{}^\alpha+\frac{1}{2\sqrt{2}}e^{-\frac{1}{2}u}
\left( \mu^{jb}\psi_{1\;bj}q_2^{i\alpha}-2\mu^{\alpha b}\psi_{1\;b\beta}q_2^{i\beta}\right),\nn \\
\delta \psi_2^{a\alpha}&=&4\im \mu_j{}^a {\dot q}{}_2^{j\alpha} +4 \im \mu_j{}^a q_2^{j\beta} J_\beta{}^\alpha+
\frac{1}{2\sqrt{2}}e^{-\frac{1}{2}u}\left( \mu^{jb}\psi_{1\;bj}\psi_2^{a\alpha} -2 \mu^{\alpha b}\psi_{1\;b\beta}\psi_2^{a\beta}\right)
\eea
if we take into account that on the reduction constraints \p{reduction2} the current $J^{\alpha\beta}$ \p{J} acquires form
\be\label{5a}
J^{\alpha\beta}=\frac{2\im}{e^u+y}\left( m K^{\alpha\beta}-\frac{1}{16}H^{\alpha\beta}-\im V^{\alpha\beta}\right),
\ee
with
\be\label{6a}
 \quad K^{\alpha\beta}=\frac{1}{2}\left( v^\alpha{\bar v}{}^\beta+ v^\beta{\bar v}{}^\alpha\right).
\ee

To construct the Hamiltonian, as usual, one should define the momenta $(p_u, p_{i\alpha}, \pi_{1\;a\alpha},
\pi_{2\;a\alpha}, p_\alpha,{\bar p}^\alpha)$ for the variables
$(u, q_2^{i\alpha}, \psi_1^{a\alpha}, \psi_2^{a\alpha}, v^\alpha ,{\bar v}_\alpha)$,
respectively
\bea\label{7a}
&&p_u=\frac{1}{2}e^u {\dot u},\quad
p_{i\alpha}=\frac{2}{e^u+y}\left( e^u{\dot q}_{2\;i\alpha}+(q_2\cdot {\dot q}_2) q_{2\;i\alpha}\right)+\frac{4\im}{e^u+y}\left(mK_{\alpha\beta}-
\frac{1}{16}H_{\alpha\beta}\right)q_{2\;i}^\beta ,\nn\\
&& \pi_{1\;a\alpha}=\frac{\im}{8}\psi_{1\;a\alpha},\quad \pi_{2\;a\alpha}=\frac{\im}{8}\psi_{2\;a\alpha},\quad
p_\alpha=\im m {\bar v}_\alpha,\quad {\bar p}{}^\alpha=-\im m v^\alpha.
\eea
Now we introduce the canonical Poisson brackets
\bea\label{8a}
&& \left\{ u,p_u\right\}=1,\quad \left\{q_2^{i\alpha},p_{j\beta}\right\}=\delta^i_j \delta^\alpha_\beta, \quad
\left\{ v^\alpha, p_\beta\right\}= - \delta^\alpha_\beta,\quad\left\{ {\bar v}_\alpha, {\bar p}{}^\beta\right\}= - \delta_\alpha^\beta, \nn\\
&& \left\{ \psi_1^{a\alpha}, \pi_{1\;b\beta} \right\}=-\delta^a_b \delta^\alpha_\beta, \quad
\left\{ \psi_2^{a\alpha}, \pi_{2\;b\beta} \right\}=-\delta^a_b \delta^\alpha_\beta.
\eea
{}From the explicit form of the fermionic $(\pi_{1\;a\alpha}, \pi_{2\;a\alpha})$ and bosonic $(p_\alpha,{\bar p}{}^\alpha)$  momenta \p{8a} it follows that we have a
second-class constraints. In order to resolve them, one has pass to the Dirac brackets for the canonical variables\footnote{From now on,
the symbol $\{,\}$ stands for the Dirac brackets.}
\bea\label{9a}
&& \left\{ u,p_u\right\}=1,\quad \left\{q_2^{i\alpha},p_{j\beta}\right\}=\delta^i_j \delta^\alpha_\beta, \quad
\left\{ v^\alpha, {\bar v}_\beta\right\}= \frac{\im}{2m}\delta^\alpha_\beta, \nn\\
&& \left\{ \psi_1^{a\alpha}, \psi_{1\;b\beta} \right\}=4\im \delta^a_b \delta^\alpha_\beta, \quad
\left\{ \psi_2^{a\alpha}, \psi_{2\;b\beta} \right\}=4\im\delta^a_b \delta^\alpha_\beta.
\eea
Let us note, that in virtue of \p{9a} the currents $K_{\alpha\beta}$ \p{6a} obey to $su(2)$ algebra
\be\label{10a}
\left\{ K^{\alpha\beta},K^{\gamma\rho} \right\} = - \frac{\im}{4m} \left( \epsilon^{\alpha\gamma}K^{\beta\rho}+
\epsilon^{\beta\rho}K^{\alpha\gamma}+
\epsilon^{\alpha\rho}K^{\beta\gamma}+
\epsilon^{\beta\gamma}K^{\alpha\rho}\right).
\ee

Finally, one may check that the following supercharges
\bea\label{11a}
{\mathbb Q}_{ia}&=&\frac{1}{\sqrt{2}}e^{-\frac{1}{2}u}p_u \psi_{1\;ai}-\frac{1}{2}p_{i\alpha}\psi_{2\;a}^\alpha-
\frac{1}{2\sqrt{2}}\left((q_2\cdot p)\psi_{1\;ai}+2\psi_{1\;a\gamma}q_{2\;k}^\gamma p^k{}_i\right)\nn\\
&& + \im \sqrt{2} e^{-\frac{1}{2}u}\left(  m K_{i j} -\frac{1}{16} H_{i j}\right) \psi_{1\;a}{}^{j}
\eea
and the Hamiltonian\footnote{Note, that in virtue of \p{6a} and \p{H} $K^{\alpha\beta}K_{\alpha\beta}=-1/2$.}
\be\label{12a}
\mathbb{H}=e^{-u}p_u^2+\frac{(q_2 \cdot p)^2}{4(e^u+y)}+\frac{e^{-u}}{4(e^u+y)} {\cal P}^{i\alpha}{\cal P}_{i\alpha}-
\frac{2}{e^u+y} \left( mK_{\alpha\beta}-\frac{1}{16}H_{\alpha\beta}\right)\left( mK^{\alpha\beta}-\frac{1}{16}H^{\alpha\beta}\right),
\ee
where
\be\label{13a}
{\cal P}_{i\alpha}=(e^u+y)p_{i\alpha}-q_{2\;i\alpha}(q_2\cdot p)-4 \im \left(m K_{\alpha\beta} -\frac{1}{16}H_{\alpha\beta}\right)q_{2\;i}{}^\beta,
\ee
form the standard \Nf superalgebra
\be\label{14a}
\left\{\mathbb{Q}{}^{ia},\mathbb{Q}_{jb}\right\}=2\im \delta^i_j \delta^a_b \mathbb{H}.
\ee

It is interesting to note that the term  with the ``coupling
constant'' $m$, which defines the interaction with the Yang
monopole, enters the supercharges $\mathbb{Q}_{ia}$ \p{11a} in a
simple way similar to ~\cite{SKOL,SBSKAS}.

Another funny peculiarity of the supercharges \p{11a} and
Hamiltonian \p{12a} is their invariance with respect to
simultaneous ``reflection''
$(q_2, p, \psi_2) \rightarrow (-q_2 ,-p, -\psi_2)$. This invariance
means that one may immediately go to the simplest case by
consistently putting $(q_2, p, \psi_2)=0$. The resulting
supercharges and Hamiltonian will describe the one-dimensional
system discussed in \cite{fil0,bk,KL1}.

With this, we completed the classical description of \Nf
five-dimensional supersymmetric mechanics describing the isospin
particle interacting with a Yang monopole. Next, we analyze some
possible extensions of the present system, together with some
possible interesting special cases.

\setcounter{equation}0
\section{Generalizations and the cases of a special interest}
In the previous Sections we have analyzed the simplest variant of $SU(2)$ reduction procedure applied to the
free eight-dimensional system with \Nf supersymmetry. Here we will consider
its possible generalizations concentrating on the bosonic sector only, while
the full supersymmetric action could be easily reconstructed, if needed.
\subsection{$SO(4)$ invariant systems}
The most general system which still possesses  $SO(4)$ symmetry upon $SU(2)$ reduction is specified by the prepotential
${\cal F}$ \p{action1} depending on two scalars $X$ and $Y$
\be\label{SP1}
{\cal F}={\cal F}(X,Y), \qquad X={\cal Q}^{i\hat\alpha}_1 {\cal Q}_{1\;i\hat\alpha},\qquad Y={\cal Q}^{i\hat\alpha}_2 {\cal Q}_{2\;i\hat\alpha}.
\ee
Such a system is invariant under $SU(2)$ transformations realized on the ``hatted'' indices $\hat\alpha$ and thus the $SU(2)$ reduction
we discussed in the Section 2 goes in the same manner. In addition the full $SU(2)\times SU(2)$ symmetry realized on the superfield
${\cal Q}^{i\hat\alpha}_2$ will survive in the reduction process. So we expected the final system will possess $SO(4)$ symmetry.

The bosonic sector of the system with prepotential \p{SP1} is described by the action
\be\label{ac1}
S=\int dt \left[ \left( F_x+\frac{1}{2}x F_{xx}\right) {\dot Q}^{i\hat\alpha}_1 {\dot Q}_{1\;i\hat\alpha}+
\left( F_y+\frac{1}{2}y F_{yy}\right) {\dot Q}^{i\hat\alpha}_2 {\dot Q}_{2\;i\hat\alpha}+  2 F_{xy} {Q}^{j\hat\beta}_2 {Q}_{1\;j\hat\alpha}{\dot Q}_{2\;i\hat\beta} {\dot Q}^{i\hat\alpha}_1\right].
\ee
Even with a such simple prepotential the bosonic action \p{ac1} after reduction has a rather complicated form. Next, still meaningful
simplification, could be achieved with the following prepotential
\be\label{SP2}
{\cal F}={\cal F}(X,Y) = {\cal F}_1 (X)+{\cal F}_2 (Y),
\ee
where ${\cal F}_1 (X)$ and ${\cal F}_2(Y)$ are arbitrary functions depending on $X$ and $Y$, respectively. With a such prepotential the third
term in the action \p{ac1} disappeared and the action acquires readable form. With our notations \p{eq1}, \p{coord} the reduced action reads
\bea\label{gen1}
S&=& \int dt \left[ \frac{H_x H_y}{2\left( (H_x-H_y) z_5 +(H_x+H_y) r\right)}{\dot z_\mu} {\dot z_\mu}
+\frac{(H_x-H_y)^2}{8 r^2 \left((H_x-H_y) z_5 +(H_x+H_y) r\right)} \left( z_\mu {\dot z_\mu}\right)^2 +\right. \nn \\
& &+\frac{H_x-H_y}{4 r^2}\left(z_\mu {\dot z_\mu}\right) {\dot z_5} +\frac{1}{8}\left( \frac{H_x-H_y}{r^2} z_5  +\frac{H_x+H_y}{r}\right){\dot z_5}^2\nn
+
im\left( {\dot v}{}^\alpha{\bar v}_\alpha -v^\alpha\dot{\bar v}_\alpha\right)\\
& &\left. -\frac{2 m^2}{(H_x+H_y)r+(H_x-H_y)z_5} -\frac{8im H_y}{(H_x+H_y)r+(H_x-H_y)z_5} v^\alpha{\bar v}{}^\beta V_{\alpha\beta} \right],
\eea
where
\be
H_x= F_{1}'(x)+\frac{1}{2} x F_{1}''(x),\qquad H_y= F_{2}'(y)+\frac{1}{2} y F_{2}''(y),
\ee
and
\be
x=\frac{1}{2}\left(r+ z_5 \right), \qquad y=\frac{1}{2}\left( r-z_5\right).
\ee
Let us stress, that the unique possibility to have $SO(5)$ invariant bosonic sector is to choose $H_x=H_y=const$. This is just the case
we considered in the Section 2. With arbitrary potentials $H_x$ and $H_y$ we have a more general system with the action \p{gen1},
describing the motion of the \Nf supersymmetric particle in five dimensions and interacting with Yang monopole and some specific potential.

\subsection{Non-linear supermultiplet}
It is known for a long time that in some special cases one could reduce the action for hypermultiplets to the action containing
one less physical bosonic components -- to the action of so-called non-linear supermultiplet \cite{NLM0,ikl1,NLM2}. The key idea of such reduction is
replacement of the time derivative of the ``radial'' bosonic component of hypermultiplet $Log(q^{ia}q_{ia})$ by an auxiliary component $B$ without breaking of \Nf supersymmetry \cite{root}:
\be\label{NLM}
\frac{d}{dt} Log(q^{ia}q_{ia})\; \rightarrow\; B.
\ee
Clearly, to perform such replacement in  some action the  ``radial'' bosonic component has to enter this action only with
time derivative. This condition is strictly constraints the variety of the possible hypermultiplet actions in which this reduction works.

To perform the reduction from hypermultiplet to the non-linear one, the parametrization \p{eq1} is not very useful. Instead, we choose
the following parameterizations for independent components of two hypermultiplets $q_1^{i\alpha}$ and  $q_2^{i\alpha}$
\be\label{eq2}
q_1^{i\alpha}=\frac{1}{\sqrt{2}}\epsilon^{i\alpha} e^{\frac{1}{2}u},\quad q_2^{i\alpha}= x^{(i\alpha)}-\frac{1}{\sqrt{2}}\epsilon^{i\alpha}z_4,
\ee
where, as  before \p{coord}
\be\label{coord2}
x^{12}=\frac{i}{\sqrt{2}}z_3, \quad x^{11}=\frac{1}{\sqrt{2}}\left( z_1+i z_2\right), \quad x^{22}=\frac{1}{\sqrt{2}}\left( z_1-i z_2\right).
\ee
Thus, the five independent components are $u$ and  $z_\mu, \mu=1,...,4$,  and
\be
x=q_1^2=e^u,\qquad y=q_2^2= \sum_{\mu=1}^4 z_\mu z_\mu \equiv r_4^2.
\ee
With this parametrization the action \p{gen1} acquires the form
\bea\label{gen2}
S&=& \int dt \left[ \frac{G_1 G_2 e^{u}}{e^u G_1+G_2 r_4^2}{\dot z_\mu} {\dot z_\mu}
+\frac{G_2^2}{e^u G_1+G_2 r_4^2} \left( z_\mu {\dot z_\mu}\right)^2 +\frac{1}{4}G_1 e^u {\dot u}{}^2\right. \nn \\
&& \left.+
im\left( {\dot v}{}^\alpha{\bar v}_\alpha -v^\alpha\dot{\bar v}_\alpha\right) -\frac{m^2}{e^u G_1+G_2 r_4^2} -\frac{4 i m G_2}{e^u G_1+G_2 r_4^2} v^\alpha{\bar v}{}^\beta V_{\alpha\beta} \right],
\eea
where
\be
G_1=G_1(u)=F_{1}'(x)+\frac{1}{2} x F_{1}''(x),\qquad G_2=G_2(r_4)=F_{2}'(y)+\frac{1}{2} y F_{2}''(y).
\ee
{}From explicit form of the action \p{gen1} follows, that if we  choose $G_1=e^{-u}$, than the ``radial'' bosonic component $u$
will enter the action only through kinetic term $\sim {\dot u}{}^2$. Thus, performing replacement \p{NLM} and excluding the auxiliary
field $B$ by its equation of motion we will finish with the action
\be\label{gen3}
S= \int dt \left[ \frac{ G_2}{1+G_2 r_4^2}\left({\dot z_\mu} {\dot z_\mu}+G_2\left( z_\mu {\dot z_\mu}\right)^2\right)
 +
im\left( {\dot v}{}^\alpha{\bar v}_\alpha -v^\alpha\dot{\bar v}_\alpha\right)
-\frac{m^2}{1+G_2 r_4^2} -\frac{4 i m G_2}{1+G_2 r_4^2} v^\alpha{\bar v}{}^\beta V_{\alpha\beta} \right].
\ee
The action \p{gen3} describes the motion of an isospin particle on four-manifold with $SO(4)$ isometry
carrying the non-Abelian field of a BPST instanton and some special potential.

\subsection{Ordinary and twisted hypermultiplets}
One more possibility to generalize the results we presented in the previous Section is to consider simultaneously ordinary hypermultiplet
${\cal Q}^{j\hat\alpha}$ obeying to  \p{con} together with twisted hypermultiplet ${\cal V}^{a\hat\alpha}$ - a quartet of
\Nf superfields subjected to constraints \cite{ikl1}
\be\label{conT}
D^{i(a}{\cal V}^{b)\hat\alpha}=0\ ,  \und
\left( {\cal V}^{a\hat\alpha}\right)^\dagger = {\cal V}_{a\hat\alpha}\ .
\ee
The most general system which is explicitly invariant under $SU(2)$ transformations realized on the ``hatted'' indices
is defined, similarly to \p{SP1},  by the superspace action depending on two scalars $X, Y$
\be\label{SPT}
S\=\int\!\!\di{t}\,\di^4\theta\ {\cal F}(X,Y), \qquad X={\cal Q}^{i\hat\alpha} {\cal Q}_{i\hat\alpha},\qquad
Y={\cal V}^{a\hat\alpha} {\cal V}_{a\hat\alpha}.
\ee
The bosonic sector of the action \p{SPT} is a rather simple
\be\label{acT}
S=\int dt \left[ \left( F_x+\frac{1}{2}x F_{xx}\right) {\dot Q}^{i\hat\alpha} {\dot Q}_{i\hat\alpha}-
\left( F_y+\frac{1}{2}y F_{yy}\right) {\dot V}^{a\hat\alpha} {\dot V}_{a\hat\alpha}\right].
\ee
Thus, we see that the term causes most complicated structure of the action with two hypermultiplets, disappeared in the case
of ordinary and twisted hypermultiplets. Clearly, the bosonic action after $SU(2)$ reduction will have the same form \p{gen1},
but with
\be\label{hh}
H_x= F_x+\frac{1}{2} x F_{xx},\qquad H_y= -\left(F_y+\frac{1}{2} y F_{yy}\right).
\ee
Here $F=F(x,y)$ is still function of two variables $x$ and $y$. The mostly symmetric situation again corresponds to the choice
\be\label{hh1}
H_x=H_y\equiv h(x,y)
\ee
with the action
\be\label{genT}
S= \int dt \left[ \frac{h}{4r}{\dot z_m} {\dot z_m}
 +
im\left( {\dot v}{}^\alpha{\bar v}_\alpha -v^\alpha\dot{\bar v}_\alpha\right)
 -\frac{ m^2}{h\; r} -\frac{4im }{r} v^\alpha{\bar v}{}^\beta V_{\alpha\beta} \right].
\ee Due to the definitions \p{hh}, \p{hh1} the metric $h(x,y)$
cannot be chosen to be fully arbitrary. For example, looking for
$SO(5)$ invariant model with $h=h(x+y)$ we could find only two
solutions \footnote {The same metric has been considered in
\cite{ONI}.} \be\label{sols} h_1 = const, \qquad h_2 = 1/(x+y)^3.
\ee Both solutions describe a cone-like geometry in the bosonic
sector.

Finally, we would like to point the attention to the fact that with $h(x,y)=const$ the bosonic sectors of the systems with two
hypermultiplets and with one ordinary and one twisted hypermultiplets are coincide. This is just one more justification that
``almost free" systems could be supersymmetrized in the different ways.

\section{Conclusion}
In the present paper, starting with the non-interacting system of
two \Nf hypermultiplets, we perform a reduction over the
$SU(2)$ group which commutes with supersymmetry. The resulting
system describes the motion of an isospin carrying particle on a
conformally flat five-dimensional manifold in the non-Abelian
field of a Yang monopole and in some scalar potential. The most
important step for this construction is passing to new bosonic and
fermionic variables, which are inert under the $SU(2)$ group, over
which we perform the reduction. Thus, the $SU(2)$ group rotates
only three bosonic components, which enter the action through
$SU(2)$ invariant currents. Just these bosonic fields become  the
``isospin'' variables, which the background field couples to. Due
to the commutativity of \Nf supersymmetry with the reduction
$SU(2)$ group, it survives upon reduction.  We also presented the
corresponding supercharges and Hamiltonian, which form a standard
\Nf superalgebra. Some possible generalizations of the action to
the cases of systems with a more general bosonic action, a
four-dimensional system which still includes eight fermionic
components, and a variant of five-dimensional \Nf mechanics
constructed with the help of the ordinary and twisted \Nf
hypermultiplets were considered. The main preference of the
proposed approach is its applicability to any system which
possesses $SU(2)$ invariance. If, in addition, this $SU(2)$
commutes with supersymmetry, then the resulting system will be
automatically supersymmetric.

One of the interesting peculiarities of the constructed system is
a very simple dependence of the supercharges on the ``coupling
constant''. Another interesting feature is the existence of two different
\Nf supersymmetrizations of the same bosonic action firstly proposed in~\cite{brazil}.

Among possible direct applications of our construction there are
the reduction in the cases of systems with non-linear \Nf
supermultiplets \cite{nelin1}, systems with more than two
(non-linear)hypermultiplets, in the systems with bigger
supersymmetry, say for example \Ne, etc. However, the most
important case, which is still missing within our approach, is the
construction of the \Nf supersymmetric particle on the sphere
$S^5$ in the field of a Yang monopole. Unfortunately, the use of
standard linear hypermultiplets makes the solution of this task
impossible because the resulting bosonic manifolds
have a different structure (conical geometry) to include $S^5$.  Thus, our hopes to
construct such a system are related either to less supersymmetric
theories (say with \Nt supersymmetry) or to non-linear
supermultiplets.

\section*{Acknowledgements}
We thank Armen Nersessian and Francesco Toppan for useful discussions.
S.K. and A.S. are grateful to the Laboratori Nazionali di Frascati for hospitality.
This work was partially supported by the grants RFBF-09-02-01209 and 09-02-91349, by
Volkswagen Foundation grant~I/84 496 as well as by the ERC Advanced
Grant no. 226455, \textit{``Supersymmetry, Quantum Gravity and Gauge Fields''%
} (\textit{SUPERFIELDS}).

\end{document}